\documentclass[twocolumn,showpacs,superscriptaddress]{revtex4} 
\usepackage{epstopdf}
\usepackage{dcolumn}   % needed for some tables
\usepackage{bm}        % for math
\usepackage{amssymb}   % for math
\usepackage{feynmp}
\usepackage{feyn}

\usepackage{verbatim}
\usepackage{mathrsfs}
\usepackage{latexsym} 
\usepackage[letterpaper]{geometry}
\usepackage[dvips]{graphicx} 
\usepackage{bbm}
\usepackage{amsmath}
\usepackage{amssymb}

\usepackage{pstricks}
\usepackage{color}

\providecommand{\tabularnewline}{\\}
\providecommand{\be}{\begin{equation}}
\providecommand{\ee}{\end{equation}}
\providecommand{\bea}{\begin{eqnarray}}
\providecommand{\eea}{\end{eqnarray}}
\providecommand{\beas}{\begin{eqnarray*}}
\providecommand{\eeas}{\end{eqnarray*}}

\providecommand{\beni}{\begin{equation*}}
\providecommand{\eeni}{\end{equation*}}

\providecommand{\bw}{\begin{widetext}}
\providecommand{\ew}{\end{widetext}}
 \providecommand{\no}{\nonumber}

\makeatother

\begin{document}
\title{A Renormalization Group computation of the critical exponents of hierarchical  spin glasses}
\author{Michele Castellana} 
\affiliation{Dipartimento di Fisica, Universit\`a di Roma `La Sapienza' , 00185 Rome, Italy}
\affiliation{LPTMS, CNRS and Universit\'{e} Paris-Sud, UMR8626, B\^{a}t. 100, 91405 Orsay, France} 
\author{Giorgio Parisi}
\affiliation{Dipartimento di Fisica, Universit\`a di Roma `La Sapienza' , 00185 Rome, Italy}
\begin{abstract} 
In a recent work (M Castellana and G Parisi, Phys. Rev. E \textbf{82}, 040105(R) (2010)), the large-scale behaviour of the simplest non-mean field spin-glass system has been analysed, and the critical exponent related to the divergence of the correlation length  computed at two loops within the $\epsilon$-expansion technique with two independent methods.
% In the first, the infrared limit of the theory is reached by a discrete coarse-graining process where the volume of the system is doubled at each step. In the latter, such limit is realized by performing the infinite-volume limit in the beginning,  removing the resulting infrared divergences by means of the ordinary Renormalization-Group techniques, and then taking the infrared limit of the renormalized field theory.  In both methods, we employed all the underlying ideas of the Renormalization Group for ferromagnetic systems, such as the existence of a unique length scale diverging at the critical point, and universality. 

%CCCCCCCCCCC
By performing the explicit calculation of the critical exponents at two loops, one obtains that the two methods yield the same result. This shows that such underlying renormalization group ideas apply  also in this disordered model, in such a way that an $\epsilon$-expansion can be consistently set up.\\
The question of the extension to  high-orders of this  $\epsilon$-expansion is  particularly interesting from the physical point of view. Indeed, once high orders of the series in $\epsilon$ for the critical exponents are known, one could check the convergence properties of the series, and find out if the ordinary series re-summation techniques yielding very accurate predictions for the Ising model  work also for this model.  If this is the case, a consistent and predictive non-mean field theory for such disordered system could be established. \\
In that regard, in this work we expose the underlying techniques of such a two-loop computation  (M Castellana and G Parisi, Phys. Rev. E \textbf{82}, 040105(R) (2010)).  We  show with an explicit example that such a computation could be quiet easily automatized, i. e. performed  by a computer program, in order to compute the $\epsilon$-expansion at high orders in $\epsilon$, and so eventually make this theory physically predictive. Moreover, all the underlying  renormalization group ideas implemented in such a computation are widely discussed and exposed. 
%CCCCCCCCCCCC
\end{abstract}
\maketitle
\section{Introduction}

The understanding of glassy systems and their critical properties is a subject of main interest in Statistical Physics. The mean-field theory of spin-glasses   \cite{MPV} and structural glasses   \cite{castellani-cavagna} provides a physically and mathematically rich theory.  Nevertheless, real Spin-Glass systems have short-range interactions, and thus cannot be successfully described by mean-field models \cite{MPV}. This is the reason why the development of a predictive and consistent theory of glassy phenomena going beyond mean field is still one of the most  hotly debated, difficult and challenging problems in this domain   \cite{6,7,8}, so that a theory describing real glassy systems is still missing. Indeed, the standard field theory techniques   \cite{zinnjustin, wilsonkogut} yielding the Ising model critical exponents with striking agreement with experimental data do not usually apply to locally-interacting glassy systems. As a matter of fact, a considerable difficulty in the set-up of a  loop-expansion for a locally-interacting Spin-Glass is that the mean-field saddle-point has a very complicated structure  \cite{parisi}, and could  be not uniquely-defined  \cite{6}. It follows that the predictions of a loop-expansion performed around one selected saddle-point could actually depend on the choice of the saddle-point itself, resulting into an intrinsic ambiguity in the physical predictions.  Moreover, non-perturbative effects are poorly understood and not under control, and the basic properties of large scale behaviour of these systems are still far from being clarified.

In ferromagnetic systems, the physical properties of the paramagnetic-ferromagnetic transition emerge in a clear way already in the original approach of Wilson   \cite{wilsonkogut}, where one can write a simple Renormalization Group (RG) transformation. It was later realized that Wilson's equations are exact in the in models with ferromagnetic power-law interactions on hierarchical lattices as the Dyson model   \cite{dyson, collet-eckmann}. Indeed, in such models one can write exact equations for the magnetization probability distribution, containing all the relevant physical informations about the paramagnetic, ferromagnetic and critical fixed point, and  the existence of a finite-temperature phase transition. In other words,  all the physical RG ideas are encoded in such recursion relations, whose solution can be  explicitly built up with the $\epsilon$-expansion technique  \cite{cassandro}. 
\\

The extension of this approach to random systems is available only in a few cases.
 
  Firstly, an RG analysis for random models on the Dyson hierarchical lattice has been pursued in the past   \cite{theumann1,theumann2}, and a systematic analysis of the physical and unphysical infrared (IR) fixed points has been developed  within the $\epsilon$-expansion technique. Unfortunately, in such models spins belonging to the same hierarchical block interact each other with the same   \cite{theumann1} random coupling $J$, in such a way that frustration turns out to be relatively weak and they are not a good representative of realistic strongly frustrated system.
%CCCCC

 Secondly, models with local interactions on hierarchical lattices built on diamond plaquettes   \cite{berker1},  have been widely studied   \cite{berker2, berker3, berker4, berker5, berker6} in their spin glass version, and lead  also to weakly frustrated systems even in their mean-field limit   \cite{gardner}. Notwithstanding this, such models yield a very useful and interesting playground to show how to implement the RG ideas in  disordered hierarchical lattices, and in particular on the construction  of  a suitable decimation rule for a frustrated system, which is one of the basic topics in the construction of a RG for  spin-glasses, and so in the identification of the existence of a spin-glass phase in finite dimension. \\ 
%CCCCCC

Moreover,  there has recently been a new wave of interest for strongly frustrated random models on  hierarchical lattices   \cite{castellana, franzjorg,XXX}: 
for example, it  has been shown   \cite{castellana} that a generalization of the Dyson model to its disordered version (the Hierarchical Random Energy Model (HREM)) has a Random Energy Model-like phase transition, yielding interesting new critical properties that don't appear in the mean-field case. \\

In a recent work  \cite{io}, we performed a field theory analysis of the critical behavior of a generalization of Dyson's model to the disordered case, known as the Hierarchical Edwards-Anderson model (HEA)  \cite{franzjorg}, that is physically more realistic than the HREM and presents a strongly-frustrated non-mean field interaction structure, being thus a good candidate to mimic the critical properties of a real spin-glass. Moreover, the symmetry properties of the HEA make an RG analysis  simple enough to be performed with two independent methods, to check if the IR-limit of the model is physically well-defined independently on the computation technique that one uses. Another element of novelty of the HEA is that its hierarchical structure makes the RG equations simple enough to make an high-order $\epsilon$-expansion tractable by means of a symbolic manipulation program, resulting in a quantitative theory for the critical exponents beyond mean field for a strongly-frustrated spin-glass system. It is possible that such a perturbative expansion turns out to be non convergent: if this happens, it may help us to pin down the non-perturbative effects. Motivated by this purpose, we have shown   \cite{io} with a two-loop calculation that such $\epsilon$-expansion can be set up consistently, and that the ordinary RG underlying ideas  actually apply also in this case, so that the IR limit of the theory is well-defined independently on the regularization technique.\\

%CCCCCCCCCCC
In the present work, we show how   the underlying RG ideas emerge in the computation, and in particular how such a  calculation has been performed, so that the reader can fully understand and reproduce it. Moreover, we show by an explicit example of such computation how this $\epsilon$-expansion could be automatized, i. e. implemented  by a computer program, and so pushed to high orders to establish its summability  properties.\medskip\\
%CCCCCCCCCCCCCCC

The HEA  is defined    \cite{franzjorg, io} as a system of $2^{k+1}$ Ising spins $S_0,\ldots,S_{2^{k+1}-1}$, with an energy function defined recursively by coupling two systems of $2^k$ Ising spins
\begin{eqnarray}\label{1}
H_{k+1}^{J}\left[S_0,\ldots,S_{2^{k+1}-1}\right] = &&\\ \nonumber
 H_{k}^{J_1}\left[S_0,\ldots,S_{2^{k}-1}\right] +  
 H_{k}^{J_2}\left[S_{2^k},\ldots,S_{2^{k+1}-1}\right]+ &&\\ \nonumber
-\frac{1}{2^{(k+1)\sigma}} \sum_{i<j}^{0,2^{k+1}-1} J_{ij} S_i  S_j ,&&
\end{eqnarray}
where 
\[
H_{1}^{J}[S_1,S_2] = -J 2^{-\sigma} S_1 S_2,
\]
and $J_{ij},\, J$ are Gaussian random variables with zero mean and unit variance.\\
 As we will show in the following,  this form of the Hamiltonian  corresponds in dividing the system in hierarchical embedded blocks of size $2^k$ and that the interaction of two spins depends on the distances of the blocks to which they belong   \cite{franzjorg,castellana}. Here $0 < \sigma < 1$ is a parameter tuning the decay of the interaction strength with distance. %CCCCCCC
The HEA is therefore a hierarchical counterpart
of the one-dimensional spin glass with power-law interactions   \cite{8} which has received
attention recently   \cite{young1, young2, young3, leuzzi, parisi2}.
%CCCCCCC 

  It is crucial to observe  \cite{franzjorg} that the sum of the squares of the interaction terms that couple the two subsystems scales with $k$ as $2^{2 k (1-\sigma)}$. Hence, for $\sigma>1/2$ the  interaction energy scales sub-extensively in the system volume,  yielding a non-mean field behavior of the model, while for $\sigma<1/2$ it grows faster than the volume, and the thermodynamic limit is not defined. The interesting region we will study is thus $\sigma \in [1/2,1]$. \smallskip \\
An equivalent definition of the HEA can be given without using the recursion relation (\ref{1}). Indeed, one can recover (\ref{1}) by defining the  HEA as a system of $2^ k$ Ising spins  with Hamiltonian
\be \label{214}
H_k[S]=-\sum_{i,j=0}^ {2^ k-1}J_{ij}S_i S_j
\ee 
where    $J_{ij}$ are Gaussian random variables with zero mean and  variance $\sigma^2_{ij}$.  The form of $\sigma^2_{ij}$ is given by the following expression: if  only the last $m$ digits in the binary representation of the points $i$ and $j$ are different,  $\sigma^2_{ij}= 2^{-2\sigma m}$. This form of the Hamiltonian  corresponds in dividing the system in hierarchical embedded blocks of size $2^m$, such that the interaction between two spins depends on the distance of the blocks to which they belong. It is important to observe that the quantity $\sigma^2_{ij}$ is not translational invariant, but it is invariant under a huge symmetry group and this will be crucial in the study of the model. \\The two definitions (\ref{1}) and (\ref{214}) are equivalent.\\ \medskip\\
 
We reproduce the IR behavior of the HEA and calculate its critical exponent by two different methods. Both methods suppose the existence of a  growing correlation scale length $\xi$, diverging for $T \rightarrow T_c$ as
\[
\xi \propto (T-T_c)^ {-\nu},
\]
in such a way that for $T\rightarrow T_c$ the theory is invariant under re-parametrizations of the length scale. \\
The first method is analogous to the coarse-graining Wilson's method for the Ising model: the scale-invariant limit is obtained by imposing invariance with respect to the composition operation of Eq. (\ref{1}), taking two systems of $2^k$ spins  and yielding a system of $2^{k+1}$ spins. As for the Dyson ferromagnetic model, thanks to the hierarchical structure of the Hamiltonian one can obtain closed formulae for  physical quantities with respect to such composition operation,  analyse the critical and non-critical fixed-points and extract $\nu$. \\
 The second method is more conventional:  the IR divergences appearing for $T\rightarrow T_c$ are removed by constructing a renormalized IR-safe theory. The fundamental physical informations one extracts from such renormalized theory are the same as those of the original theory defined by Eq. (\ref{1}). In particular, the correlation length and its power-law behavior close to the critical point must be the same, and so the critical exponent $\nu$. \medskip\\
 
 The rest of this paper is divided into three main Sections: in Sec. \ref{wilson} we go through the main steps of the  computation  with Wilson's method, 
%CCCC 
 show that the tensorial operations can be easily implemented diagrammatically, and thus performed by a computer program to extend such $\epsilon$-expansion to high orders.
 %CCC
  Moreover, we give the two-loop result for $\nu$. In Sec. \ref{zinn} the same result is reproduced with the field-theoretical method, and the  analogies between the two methods are discussed.
%CCC  
 In particular, we discuss why Wilson's method would be definitely better for an automatization of the $\epsilon$-expansion to high orders.
%CCCC  
   Both in Sections \ref{wilson} and \ref{zinn}, we explicitly do all  the steps of the calculation at one loop, giving to the reader all the information needed to reproduce the two-loop result for $\nu$. 
  
   Finally, in Sec. \ref{conc} the two-loop  result  is discussed in the perspective of the set up of a high-order $\epsilon$-expansion.

\section{Wilson's method}\label{wilson}

 As mentioned before, the hierarchical symmetry structure of the model makes the implementation of a recursion-like RG equation simple enough to be solved  within an approximation scheme. As a matter of fact, let us define the probability distribution of the overlap   \cite{MPV,castellani-cavagna} 
 \bea \label{overlap}
 &Q_{ab}, \, a=1,\cdots,n,\\ \no 
  &Q_{ab}=Q_{ba},\, Q_{aa}=0\, \forall a,b=1,\ldots, n
\eea
 as
 \bw
 \be 
 \mathcal{Z}_k[Q] \equiv \mathbb{E}_{J} \left[ \sum_{\{S_i\}_i} \exp\left( -\beta \sum_{a=1}^ n H_{k}^ J[S_0^ a,\cdots, S_{2^ k-1}^ a ] \right) \prod_{a<b=1}^ n \delta\left(Q_{ab} - \frac{1}{2^ k}\sum_{i=0}^ {2^{k}-1} S_i^ a S_i^b \right)\right],
 \ee
 \ew
 and the rescaled  overlap-distributions as 
 \[
 \mathcal{\mathcal{Z}}_k[Q]\equiv \mathcal{Z}_k[2^ {-k(1-\sigma)}Q],
 \]
 where $\beta \equiv 1/T$ is the inverse temperature.\\
 According to the general prescriptions of the replica approach, all the Physics of the model is encoded in the $n\rightarrow 0$ limit of $\mathcal{Z}_k[Q]$. \\
 
 It is easy to show that the recursion relation (\ref{1}) for the Hamiltonian results into a recursion relation for $\mathcal{\mathcal{Z}}_k[Q]$ 
\begin{eqnarray}\label{2}
&&\mathcal{\mathcal{Z}}_k[Q]=\exp\left({\frac{\beta^2}{4}  \text{Tr}\left[ Q^2 \right]} \right)\times 
\\ \no && \times \int \left[ d P \right] \mathcal{\mathcal{Z}}_{k-1}\left[ \frac{Q+P}{C^ {1/2}} \right]
\times \mathcal{Z}_{k-1}\left[ \frac{Q-P}{C^ {1/2}} \right], 
\end{eqnarray}
where $\text{Tr}$ denotes the trace over the replica indexes,   $\int \left[ d P \right]$ stands for the functional integral over $P_{ab}$ and \be \label{120}
C \equiv 2^ {2(1-\sigma)}.
\ee
 Eq. (\ref{2}) physically represents a recursion equation relating the probability distribution $\mathcal{Z}_{ k-1}[Q]$ to $\mathcal{Z}_{k}[Q]$, obtained by $\mathcal{Z}_{k-1}[Q]$ by a coarse-graining RG step, composing two subsystem of size $2^k$ to form a system of size $2^{k+1}$. Eq. (\ref{2}) is analogous to the recursion equation  in Dyson's model  \cite{cassandro}, relating the  probability distribution  $g_k(m)$ of the magnetization at the $k$-th hierarchical level  to $g_{k-1}(m)$. \bigskip \\

To illustrate the technique used to solve perturbatively (\ref{2}) for $\mathcal{Z}_k[Q]$, we will show our method in a simple toy example, where the matrix field $Q_{ab}$ is replaced by a one-component field $\phi$, the functional $\mathcal{Z}_k[Q]$ by a function $\Omega_{k}(\phi)$, and Eq. (\ref{2}) by 
\begin{eqnarray}\label{2bis}
&& \Omega_k(\phi)=\exp\left({\frac{\beta^2}{4}  \phi ^2 } \right)\times 
\\ \no && \times \int \ d \chi \,    \Omega_{k-1}\left( \frac{\phi+\chi}{C^ {1/2}} \right)
\times \Omega_{k-1}\left( \frac{\phi-\chi }{C^ {1/2}} \right).
\end{eqnarray}

  As for Dyson's model,  (\ref{2bis}) can be solved by making an ansatz for  $\Omega_{k}(\phi)$. The simplest form one can suppose for $\Omega_k(\phi)$ is the  Gaussian one
\be \label{gaussbis}
\Omega _k(\phi)=\exp \left[- (d_k \phi ^2+e_k \phi)\right]. 
\ee
This form corresponds to a mean-field solution  \cite{cassandro}. By inserting Eq. (\ref{gaussbis}) into  Eq. (\ref{2bis}), one finds a recursion equation relating $d_{k},e_{k}$ to $d_{k-1},e_{k-1}$
\beas
d_{k}&=&\frac{2 d_{k-1}}{C}- \frac{\beta^2}{4},\\
e_k & = & \frac{2 e_{k-1}}{C^ {1/2}}.
\eeas
Non-gaussian  solutions can be explicitly constructed perturbatively. Indeed, by setting  
\begin{equation}\label{3bis}
\mathcal{Z}_k[Q]=\exp\left[ - \left(d_k   \phi^2 + e_k \phi   + \frac{u_k}{3}  \phi ^3  \right) \right], 
\end{equation}
and supposing that $u_k$ is small, one can plug Eq. (\ref{3bis}) into Eq. (\ref{2bis}) and get 
\bw
\bea \label{100bis}
\Omega_k(\phi)&=&
\exp\left\{ -\left[ {\left(  \frac{2 d_{k-1}}{C}  -\frac{\beta^2}{4}  \right)\phi ^2} + \frac{2 e_k}{C^ {1/2}} \phi +\frac{2 u_{k-1}}{3 C^{3/2}}  \phi ^3 \right] \right\} \int  d \chi   \exp\left[ -\left(  \frac{2 d_{k-1}}{C}   \chi^2  + \frac{2 u_{k-1}}{ C^{3/2}}  \phi  \chi ^ 2 \right) \right]\\ \no 
&\propto &
\exp\left\{ -\left[ {\left(  \frac{2 d_{k-1}}{C}  -\frac{\beta^2}{4}  \right)\phi ^2} + \frac{2 e_k}{C^ {1/2}} \phi + \frac{2 u_{k-1}}{3 C^{3/2}}  \phi ^3 \right] \right\}    \left(  \frac{2 d_{k-1}}{C}     + \frac{2 u_{k-1}}{ C^{3/2}}  \phi  \right) ^ {-1/2}\\ \no 
& =  &
\exp\left\{ -\left[ {\left(  \frac{2 d_{k-1}}{C}  -\frac{\beta^2}{4}  \right)\phi ^2} + \frac{2 e_k}{C^ {1/2}} \phi + \frac{2 u_{k-1}}{3 C^{3/2}}  \phi ^3 + \frac{1}{2} \log \left(  \frac{2 d_{k-1}}{C}     + \frac{2 u_{k-1}}{ C^{3/2}}  \phi  \right)  \right] \right\}   \\ \no 
& \propto  &
\exp\left\{ -\left[ {\left(  \frac{2 d_{k-1}}{C}  -\frac{\beta^2}{4}  \right)\phi ^2} + \frac{2 e_k}{C^ {1/2}} \phi + \frac{2 u_{k-1}}{3 C^{3/2}}  \phi ^3 + \frac{1}{2} \log \left( 1    + \frac{ u_{k-1}}{C^{1/2} d_{k-1}}  \phi  \right)  \right] \right\}    \\ \no 
& \propto  &
\exp\Bigg\{ -\Bigg[ {\left(  \frac{2 d_{k-1}}{C}  -\frac{\beta^2}{4}  - \frac{1}{4}\left(\frac{ u_{k-1}}{C^{1/2} d_{k-1}}\right)^2 \right)\phi ^2} +  \left(\frac{2 e_{k-1}}{C^ {1/2}}+ \frac{ u_{k-1}}{2 C^{1/2} d_{k-1}}  \right)  \phi +\\ \no 
&& + \frac{1}{3} \Bigg( \frac{2 u_{k-1}}{ C^{3/2}} +\frac{1}{2}\left( \frac{ u_{k-1}}{C^{1/2} d_{k-1}}\right)^3 \Bigg)   \phi ^3  + O(u_{k-1}^ 4)\Bigg] \Bigg\}   , 
\eea
\ew
where $\propto$ stands for a $\phi$-independent proportionality constant, that will be  omitted in the following. Comparing Eq. (\ref{100bis}) with Eq. (\ref{3bis}), one finds three recurrence equations relating $d_k,e_k,u_k$ to $d_{k-1},e_{k-1},u_{k-1}$
\bea \label{105bis}
d_{k}&=& \frac{2 d_{k-1}}{C} -\frac{\beta^ 2}{4}-  \frac{1}{4 } \left( \frac{u_{k-1}}{2 C^ {1/2} d_{k-1}} \right)^2 + \\ \no 
&&+O(u_{k-1}^4),
\eea
\beas
e_k & = & \frac{2 e_{k-1}}{C^ {1/2}}+ \frac{ u_{k-1}}{2 C^{1/2} d_{k-1}} + O(u_{k-1}^3),\\ \no
u_{k}&=& \frac{ 2 u_{k-1} }{C^ {3/2}}+ \frac{1}{2}\left(\frac{u_{k-1}}{2 C^ {1/2} d_{k-1}} \right)^3+O(u_{k-1}^ 5).
\eeas
One can easily analyse the fixed points of the RG-flow equations (\ref{105bis}), and the resulting critical properties. We will not enter into these details for the toy model, since all these calculations will be illustrated extensively in the original theory. \bigskip\\

  Back to the original problem,  Eq. (\ref{2}) can be solved by making an ansatz for   $\mathcal{Z}_{k}[Q]$, following the same lines as in the toy model case. The simplest form one can suppose for $\mathcal{Z} _k[Q]$ is the  Gaussian one
\be \label{gauss}
\mathcal{Z}_k[Q]=\exp \left(- r_k \text{Tr}[Q^2]\right). 
\ee
This form corresponds to a mean-field solution   \cite{parisi,6}. By inserting Eq. (\ref{gauss}) into Eq. (\ref{2}),
one finds the evolution equation relating $r_{k-1}$ to $r_k$
\be 
r_{k}=\frac{2 r_{k-1}}{C}- \frac{\beta^2}{4}. 
\ee
\bigskip\\
Corrections to the mean-field solution can be investigated by adding  non-Gaussian terms to Eq.  (\ref{gauss}), that are proportional to  higher powers of $Q$, and consistent with the symmetry properties of the model. It is easy to see  \cite{MPV} that the only cubic term in $Q$ consistent with such symmetry conditions is $\text{Tr}[Q^3]$, so that the non-mean field ansatz of $\mathcal{Z}$ reads
\begin{equation}\label{3}
\mathcal{Z}_k[Q]=\exp\left[ - \left(r_k  \text{Tr}\left[ Q^2 \right] + \frac{w_k}{3}  \text{Tr}[Q^3]  \right) \right].
\end{equation}
This correction can be handled by supposing that the term $w_k$ proportional to the non-quadratic term is small for every $k$, and performing a systematic expansion in powers of it. By inserting Eq. (\ref{3}) into the recursion relation Eq.  (\ref{2}), one finds
\bw
\bea\label{100}
\mathcal{\mathcal{Z}}_k[Q]&=&
\exp\left\{ -\left[ {\left(  \frac{2 r_{k-1}}{C}  -\frac{\beta^2}{4}  \right)\text{Tr}\left[ Q^2 \right]} + \frac{2 w_{k-1}}{3 C^{3/2}}  \text{Tr}[Q^3] \right] \right\} \int \left[ d P \right] \exp\left[ -S^{(3)}_{k-1}[P,Q]\right],\\ \no 
S^{(3)}_{k-1}[P,Q] & \equiv & \frac{2 r_{k-1}}{C}  \text{Tr}\left[ P^2 \right] + \frac{2 w_{k-1}}{ C^{3/2}}  \text{Tr}[Q P^ 2]. 
\eea
\ew
The Gaussian integral in Eq. (\ref{100}) can be computed exactly. Indeed, setting $A \equiv (a,b) \, a>b $ and 
%\bw
\bea
\frac{\partial ^ 2 S^{(3)}_{k-1}[P,Q]  }{\partial P_A \partial P_B} & \equiv & \\ \no 
  \frac{4 r_{k-1}}{C}\delta_{AB} + \frac{ 2 w_{k-1}}{C^ {3/2}}M_{AB}[Q],&&
  \eea
  \bea  
M_{ab, cd}[Q] & \equiv & N_{ab, cd}[Q]  + N_{ab, dc}[Q] , \label{210}\\
N_{ab, cd}[Q] & \equiv & \delta_{bc} Q_{da} + \delta_{ac} Q_{db} , \label{211}
\eea
%\ew
one finds
\bw
\be \label{a}
\mathcal{\mathcal{Z}}_k[Q]=
\exp\left\{ -\left[ {\left(  \frac{2 r_{k-1}}{C}  -\frac{\beta^2}{4}  \right)\text{Tr}\left[ Q^2 \right]} + \frac{2 w_{k-1}}{3 C^{3/2}}  \text{Tr}[Q^3] \right] \right\}
  \left[ \det  \left(      \frac{4 r_{k-1}}{C}\delta_{AB} + \frac{ 2 w_{k-1}}{C^ {3/2}}M_{AB}[Q]  \right) \right]^ {-\frac{1}{2}} .  
\ee \label{aa}
\ew
The determinant in the right-hand side of Eq. (\ref{100}) can now be expanded in $w_{k-1}$. Denoting by $\textbf{Tr}$ the trace over the $A$-type indexes,
% using the standard relation $\log \det = \textbf{Tr} \log$ and 
one has to explicitly evaluate the  traces $\textbf{Tr}[M[Q]^2], \, \textbf{Tr}[M[Q]^3]$. Here we show how the  trace $\textbf{Tr}[M[Q]^2]$ can be evaluated, in order to show to the reader how the tensorial operations over the replica indexes can be generally carried out. \\
By using Eqs. (\ref{210}), (\ref{211}), one has 
\bw
\bea \label{212}
\textbf{Tr}[M[Q]^2]&=& \sum_{AB} M[Q]_{AB} M[Q]_{BA}\\ \no 
&=&\sum_{a>b, c>d} \left( N_{ab, cd}[Q]  + N_{ab, dc}[Q]\right) \times \\ \no 
&&\times \left( N_{cd, ab}[Q]  + N_{dc, ab}[Q]\right)\\  \no 
&=&\sum_{a\neq b, c\neq d}  N_{ab, cd}[Q]   N_{cd, ab}[Q] \\  \no 
&=&\sum_{a\neq b, c\neq d}  (\delta_{bc} Q_{da} + \delta_{ac} Q_{db} )\times \\  \no 
&&\times( \delta_{da} Q_{bc} + \delta_{ca} Q_{bd} ) \\  \no 
&=&\sum_{a\neq b, c\neq d}   \delta_{ca} Q_{bd}^2  \\  \no 
&=&\sum_{a bcd} (1- \delta_{ab})(1-\delta_{cd})   \delta_{ca} Q_{bd}^2  \\ \no 
&=&(n-2) \sum_{ab} Q^ 2_{ab}\\  \no 
&=&(n-2) \text{Tr}[Q^2].   
\eea
\ew
The steps in Eq. (\ref{212}) can be summarized as follows: in the second line we write the tensorial operations over the super indexes $A,B,\cdots $ in terms of the replica indexes $a,b,\cdots$, in the third line we use the symmetries of $M_{ab,cd}[Q]$ with respect to $a \leftrightarrow b$ and $c \leftrightarrow d$ and re-write the sum over $a>b,\, c>d$ in terms of a sum with $a\neq b,\, c\neq d$. In the fifth line we find out which amongst the terms stemming from  the product $(\delta_{bc} Q_{da} + \delta_{ac} Q_{db} ) ( \delta_{da} Q_{bc} + \delta_{ca} Q_{bd} ) $   vanish because of the constraints $a\neq b,\, c\neq d, \, Q_{aa}=0$, and because of the Kronecker $\delta$s in the sum. Once we are left with the non-vanishing terms, in the sixth line we write explicitly the sum over $a\neq b,\, c\neq d$ in terms of an unconstrained sum over $a,b,c,d$ by adding the constraints $(1-\delta_{ab})(1-\delta_{ cd})$. In the seventh line we perform explicitly the sum over the replica indexes, and write everything in terms of the invariant $I^{(1)}_2[Q] = \text{Tr}[Q^2]$ (see Table \ref{tab1}).\medskip\\

\begin{centering}

\begin{figure}[h]
\begin{centering}
\includegraphics[scale=0.8]{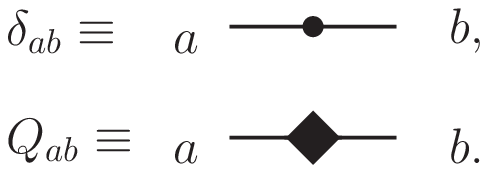}
\par\end{centering}

\caption{Graphical identifications  representing  symbolically mathematical objects used in tensorial operations. The basic objects of the tensorial operations are the $\delta_{ab}$ function imposing that the replica indexes $a$ and $b$ are equal (top), and the matrix $Q_{ab}$ (bottom). Once these elements are represented graphically, all the tensorial operations can be carried out by manipulating graphical objects composed by these elementary  objects.}
\label{identifications}  
\end{figure} 
\end{centering}

%CCCCCCCCCCCC
Here we show how the trace in  Eq. (\ref{212}) can be computed with a purely graphical method, that can be easily implemented in a computer program to perform this computation at  high orders in $w_k$. Let us set 
\bea \label{tr2}
 \textbf{Tr}_2[ f ] &\equiv &\sum_{a_1\neq b_1,\cdots,a_k\neq b_k}f_{a_1b_1,\cdots,a_k b_k}\\ \no 
 & = & \sum_{a_1 b_1,\cdots, a_k b_k}(1-\delta_{a_1 b_1}) \cdots    (1-\delta_{a_k b_k}) \times\\ \no
 && \times f_{a_1b_1,\cdots,a_k b_k}.
\eea
and make the  graphical identifications shown in Fig. \ref{identifications}.\\

\bw

\begin{centering}

\begin{figure}[p]
\begin{centering}
\includegraphics[scale=0.7]{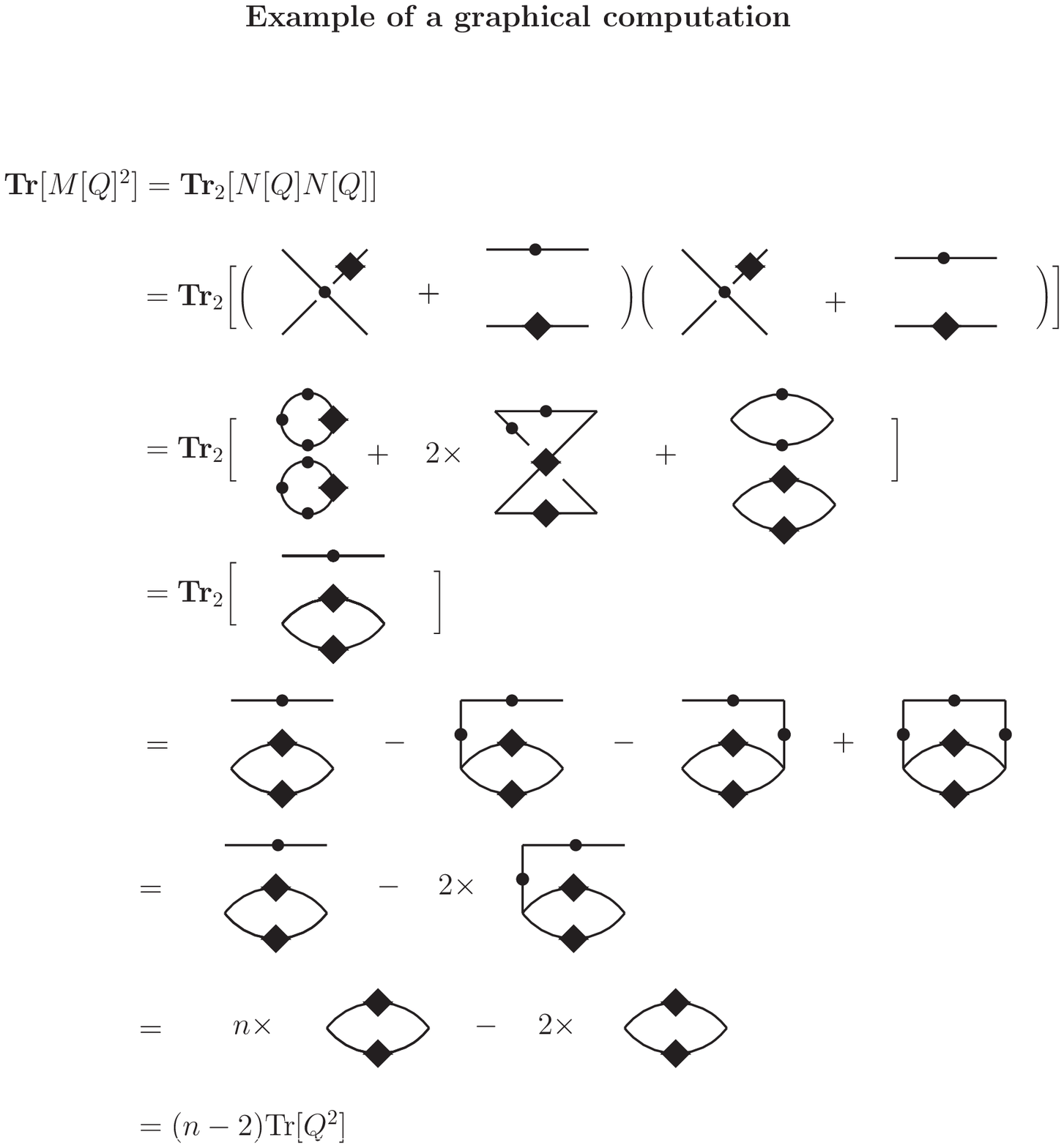}
\par\end{centering}
\caption{Graphical computation of $\textbf{Tr}[M[Q]^2]$ in Eq. (\ref{212}). \\
In the second line, the two addends of the matrix $N[Q]_{ab,cd}$  in Eq. (\ref{211}) are represented graphically in terms of the graphical objects defined in Fig. \ref{identifications}.\\
In the third line, the legs of such addends are contracted with each other, and four terms are generated. Two of them are topologically identical, hence we simply put a factor $2$ multiplying the second term in the second line.\\
 According to the condition $Q_{aa}=0$ in Eq. (\ref{overlap}), the first and second term in the third line vanish. Indeed, in these terms the lines coming out of the square vertex ($Q_{ab}$) are connected by a circuit, meaning that the matrix element $Q_{ab}$ is computed with $a=b$, and thus vanishes. Moreover, the two top-lines in the third term are actually equivalent to just one line, because of the relation $\delta_{ab}^2=\delta_{ab}$. Hence, we are left with a single term in the fourth line.\\
 In the fifth line, we perform graphically the operation $\textbf{Tr}_2$. Such an operation can be easily implemented graphically by looking at the second line of Eq. (\ref{tr2}). Let us  expand the product of $\delta$s in the second line of Eq. (\ref{tr2}), and  recall from  Fig. \ref{identifications} that $\delta_{ab}$ represents a line with a circular dot connecting 
$a$ with $b$. Hence, given graphical object $O$, with external- legs (indexes) $(a_1,b_1),\cdots, (a_k,b_k)$,  $\textbf{Tr}_2[O]$   is nothing but a sum of all the possible $2^ k$ contractions (performed with a line with a circular dot) of such external legs, where each  contracted term is multiplied by $(-1)^ {\#\text{ of contractions of the term}}$. In this case $k=2$, so we generate $2^2$ terms in the fifth line.\\
 In the sixth line, we take into account the fact that the second and third term in the fifth line are topologically isomorph, and that the fourth term in the fifth line vanishes because of the relation $Q_{aa}=0$. \\
In the seventh line the unconstrained sum over the replica indexes is finally performed. This can be done graphically in the following way: when we have an external line  with a round vertex, summing over the replica index  represented by that line means  that one has simply to remove the line (this is the graphical implementation of the relation $\sum_{b}\delta_{ab} g_b =g_a$). We do this in the first term: we sum over the top-left index, and remove the line on the top. Then we sum over the top right index by simply multiplying by $n$. The sum over the bottom-left and bottom-right indexes simply yields $\textrm{Tr}[Q^2]$. We to the same for the second term: we sum over the top-right index and remove the top line, then sum over the top-left index and remove the top-left line. Then, the sum over the bottom-left and bottom-right indexes yields $\textrm{Tr}[Q^2]$.\\
 Hence, we get the same result as in Eq. (\ref{212}).}
\label{graphcomp}  
\end{figure}
\end{centering}

\ew
The last line in Eq. (\ref{212}) can be reproduced by a purely graphical computation, as shown in Fig. \ref{graphcomp}. There we show that all the tensorial operations  have a graphical interpretation, and so that they can be performed without using the  cumbersome notation of Eq. (\ref{212}). This graphical notation is suitable for an implementation in a computer program, that could push our calculation to very high orders. For example, as shown in Fig. \ref{graphcomp} in a simple example,  while computing $\textbf{Tr}[M[Q]^k]$ for $k\gg 1$, a proliferation of terms occurs, and  some of these terms can be shown to be equal to each other, because represented by isomorph graphs, so that  the calculations can be extremely simplified. For example, in a computer program implementation of such a computation, one could identify the isomorph graphs by using some powerful graph isomorphism algorithm   \cite{alignment}. \medskip \\
%CCCCCCCCCCCCCCCCCCCC
%

%
By following these steps shown Eq. (\ref{212})
%CCCCCC
 (or their graphical implementation),
 %CCCCCCCCCCC
 all the other tensorial operations can be carried out. In particular, one finds  
\bea \label{213}
 \textbf{Tr}[M[Q]^3]&=& (n-2) \text{Tr}[Q^3].
\eea\smallskip\\
By plugging Eqs. (\ref{212}), (\ref{213}) into Eq. (\ref{a}), one finds
\bw
\bea\label{4}
\mathcal{Z}_k[Q]&=& \exp \Bigg\{ -\Bigg[ \left( \frac{2 r_{k-1}}{C} - \frac{\beta^2}{4}-\frac{n-2}{4}\left( \frac{w_{k-1}}{2 r_{k-1} C^{1/2}}\right)^2   \right) \text{Tr}[Q^2] + \\ \nonumber
&&+\frac{1}{3} \left( \frac{2 w_{k-1}}{C^ {3/2}} +\frac{n-2}{2}\left( \frac{w_{k-1}}{2 C^ {1/2}}\right)^3   \right) \text{Tr}[Q^3] + O(w_{k-1}^4)    \Bigg] \Bigg\}.
\eea
\ew
Comparing Eq. (\ref{4}) with Eq. (\ref{3}), one finds a recursion relation for the coefficients $r_k,\, w_k$
\bw
\bea \label{105}
r_{k}&=& \frac{2 r_{k-1}}{C} -\frac{\beta^ 2}{4}-  \frac{n-2}{4 } \left( \frac{w_{k-1}}{2 C^ {1/2} r_{k-1}} \right)^2 + O(w_{k-1}^4),\\ \no
w_{k}&=& \frac{ 2 w_{k-1} }{C^ {3/2}}+ \frac{n-2}{2}\left(\frac{w_{k-1}}{2 C^ {1/2} r_{k-1}} \right)^3+O(w_{k-1}^ 5).
\eea
\ew
%where $\epsilon\equiv \sigma-2/3$. \\
Eq. (\ref{5}) shows that for $\epsilon\equiv \sigma - 2/3 <0$ $w_k\rightarrow 0$ for $k\rightarrow \infty$, i. e. the corrections to the mean-field $\mathcal Z_k[Q]$ vanish in the IR limit. In this case,  the critical fixed point $(r_\ast,w_\ast)$ of Eqs. (\ref{6}), (\ref{5}) has $w_\ast=0$. On the contrary, for $\epsilon>0$ a non-trivial critical fixed-point $w_\ast\neq 0$ arises. According to  general RG arguments, such non-trivial fixed point will be proportional to $\epsilon$   \cite{zinnjustin}. In particular, one finds that $w_\ast^2 = O(\epsilon)$. \\

The critical exponent $\nu$ can be computed  \cite{wilsonkogut} by considering the $2 \times 2$ matrix $\mathcal M$ linearising the transformation given by Eq. (\ref{105}) around the critical fixed-point $(r_\ast,w_\ast)$
\[\left(
\begin{array}{c}
r_k - r_\ast\\
w_k-w_\ast
\end{array}\right) = \mathcal{M} \cdot \left(\begin{array}{c}
r_{k-1} - r_\ast\\
w_{k-1}-w_\ast
\end{array}\right),
\]
and is given by 
\be \label{80}
\nu= \frac{\log 2}{\log \Lambda},
\ee
where $\Lambda$ is the largest eigenvalue of $\mathcal{M}$.\medskip \\
%CCCCCCCCCCCC
\bw

\ew
%
%CCCCCCCCCCCCC 

Such procedure can be  systematically  pushed to higher order in $w_k$, and thus in $\epsilon$, by taking into account further corrections to the mean-field solution. Indeed, if we go back to Eq. (\ref{a}) and consider  also the $O(w_{k-1}^4)$ terms in the right side, we find  
\bw
\bea \label{102}
\left[ \det \left( \frac{4 r_{k-1}}{C}\delta_{AB} + \frac{ 2 w_{k-1}}{C^ {3/2}}M_{AB}[Q]  \right) \right]^ {-\frac{1}{2}}  & = &  \\ \no 
 \exp \Bigg\{ - \frac{1}{2} \textbf{Tr}\Bigg[  - \frac{1}{2}\left( \frac{w_{k-1}}{2 C^{1/2} r_{k-1}} \right) ^2 M[Q]^2        + \frac{1}{3}\left( \frac{w_{k-1}}{2 C^{1/2} r_{k-1}} \right) ^3 M[Q]^3   + &&\\ \no 
 - \frac{1}{4}\left( \frac{w_{k-1}}{2 C^{1/2} r_{k-1}} \right) ^4 M[Q]^4 + O(w_{k-1}^5)  \Bigg] \Bigg\}.&&
\eea
\ew
By computing explicitly the  $O(w_{k-1}^4)$ term in the right side of Eq. (\ref{102}), one finds

\bea \label{103}
\textbf{Tr}[M[Q]^4] & = &n I^{(4)}_1[Q] + 3 I^{(4)}_2[Q] +\\ \no 
&&- 16 I^ {(4)}_3[Q] - 8 I^{(4)}_4[Q]. \\ \no 	
 I^{(4)}_1[Q] &\equiv & \text{Tr}[Q^4], \\ \no 	
 I^{(4)}_2[Q] &\equiv &  ( \text{Tr}[Q^2] )^2 , \\ \no 	
 I^{(4)}_3[Q] &\equiv & \sum_{b\neq c}Q^2_{ab} Q^ 2_{ac}, \\ \no 	
 I^{(4)}_4[Q] &\equiv &  \sum_{ab}Q^4_{ab}.
\eea

Plugging Eq. (\ref{103}) in Eq. (\ref{102}) and  Eq. (\ref{102}) in Eq. (\ref{a}), we see that 
 at $O(w_{k-1}^4)$, Eq. (\ref{2})  generates  the  fourth-order monomials $\{ I^ {(4)}_l[Q] \}_{l=1,\ldots, 4}$, that are not included into the original ansatz (\ref{3}). It follows  that at  $O(w_k^4)$, $\mathcal{Z}_{k}[Q]$ must be of the form 
\bw
\begin{equation}\label{eq7}
\mathcal{Z}_k[Q]=\exp\left[ - \left(r_k  \text{Tr}\left[ Q^2 \right] + \frac{w_k}{3}  \text{Tr}[Q^3] + \frac{1}{4}\sum_{l=1}^ 4 \lambda_{l\, k} I^{(4)}_l[Q] \right) \right],
\end{equation}
\ew
with $\lambda_{l\, k}=O(w_k^4) \, \forall l=1,\ldots,4$.\medskip \\
By inserting Eq. (\ref{eq7}) into Eq. (\ref{2}) and expanding up to $O(w_{k-1}^4)$, we obtain six recursion equations relating $r_k,w_k,\lambda_{1\,k},\cdots, \lambda_{4\, k}$ to $r_{k-1},w_{k-1},\lambda_{1\, k-1},\cdots, \lambda_{4\, k-1}$. 
\bigskip\\
Such systematic  expansion can be iterated to any order $O(w_k^ p)$, obtaining 
\begin{equation}
\mathcal{Z}_k[Q]=\exp{\left( - \sum_{j=2}^{p} \sum_{l=1}^ {n_j} c^{(j)}_{l,\, k} I_l^ {(j)}[Q]  \right)},
\end{equation}
where $c^{(2)}_{1\, k}=r_k, \, c^{(3)}_{1\, k}=w_k/3,  c^{(4)}_{l\, k}=\lambda_{l\,k}/4\, \forall l=1,\ldots,4$. In this way, a recursion equation relating  $\{c^{(j)}_{l,\, k-1}\}_{j,l}$ to $ \{c^{(j)}_{l,\, k}\}_{j,l}$ is obtained. \medskip\\
The number $n_j$ of monomials generated at the step $j$ of this procedure proliferates for increasing $j$. In Table \ref{tab1} we show the  invariants $I^ {(j)}_l[Q]$ obtained by performing this systematic expansion up to the order $p=5$. It is interesting to observe that  the invariants $\text{Tr}[Q^2]^2,\,\text{Tr}[Q^2]\text{Tr}[Q^3]$ that are  generated, are of $O(n^2)$ if the matrix $Q_{ab}$ is replica symmetric. Notwithstanding this, in general they will  give a non-vanishing contribution  to the recursion relations $\{c^{j}_{l,\, k-1}\}_{j,l} \rightarrow \{c^{j}_{l,\, k}\}_{j,l}$, and so to $\nu$.  The recurrence equations at this order are the following
\bw
\bea
c^{(2)}_{1 \, k}&=& \frac{2 c^{(2)}_{1 \, k-1}}{C}-\frac{\beta^ 2}{4}-\frac{n-2}{4}\left(\frac{c^{(3)}_{1 \, k-1} }{2 C^ {1/2} c^{(2)}_{1 \, k-1}} \right)^2+ (2n-1)  \frac{c^{(4)}_{1 \, k-1}}{8 C c^{(2)}_{1 \, k-1}}+\label{220}\\  \no 
&&+  \frac{c^{(4)}_{2 \, k-1}}{2 C c^{(2)}_{1 \, k-1}}\left[ 1 + \frac{n(n-1)}{4} \right]+(n-2) \frac{c^{(4)}_{3 \, k-1}}{8 C c^{(2)}_{1 \, k-1}} + 
 \frac{3 c^{(4)}_{4 \, k-1}}{8 C c^{(2)}_{1 \, k-1}}+  O\left((c^{(3)}_{1 \, k-1} )^ 6\right), \\
c^{(3)}_{1 \, k} & = & \frac{2 c^{(2)}_{1 \, k-1}}{C^ {3/2}} + \frac{n-2}{2}\left(\frac{c^{(3)}_{1 \, k-1}}{2 C^ {1/2} c^{(2)}_{1 \, k-1}} \right)^3 + \frac{3 n c^{(5)}_{1 \, k-1}}{4 C^{3/2} c^{(2)}_{1 \, k-1}}+ (n+3)\frac{3 c^{(5)}_{2 \, k-1}}{20 C^ {3/2} c^{(2)}_{1 \, k-1}} + \frac{9 c^{(5)}_{3 \, k-1}}{20 C^ {3/2} c^{(2)}_{1 \, k-1}}+ \\    \no 
& & + \frac{3 c^{(5)}_{4 \, k-1} }{20 C^ {3/2} c^{(2)}_{1 \, k-1}}[12 + n(n-1)]- \frac{3 c^{(3)}_{1 \, k-1}}{4 C^ {1/2} c^{(2)}_{1 \, k-1}} \left[  \frac{(n-1) c^{(4)}_{1 \, k-1}}{2 C c^{(2)}_{1 \, k-1}}  +  \frac{2 c^{(4)}_{2 \, k-1}}{2 C c^{(2)}_{1 \, k-1}} +  \frac{c^{(4)}_{3 \, k-1}}{2 C c^{(2)}_{1 \, k-1}} \right]+\\ \no 
&&+ O\left((c^{(3)}_{1 \, k-1} )^ 6\right),\\   
c^{(4)}_{1 \, k} & = & \frac{2 c^{(4)}_{1 \, k-1}}{C^2}- \frac{n}{2}\left(\frac{ c^{(3)}_{1 \, k-1} }{2 C^ {1/2} c^{(2)}_{1 \, k-1}} \right)^4+ O\left((c^{(3)}_{1 \, k-1} )^ 6\right),   \label{110}\\
c^{(4)}_{2 \, k} & = & \frac{2 c^{(4)}_{2 \, k-1}}{C^2}- \frac{3}{2}\left(\frac{ c^{(3)}_{1 \, k-1} }{2 C^ {1/2} c^{(2)}_{1 \, k-1}} \right)^4+ O\left((c^{(3)}_{1 \, k-1} )^ 6\right),\\   
c^{(4)}_{3 \, k} & = & \frac{2 c^{(4)}_{3 \, k-1}}{C^2}+ 8\left(\frac{ c^{(3)}_{1 \, k-1} }{2 C^ {1/2} c^{(2)}_{1 \, k-1}} \right)^4+ O\left((c^{(3)}_{1 \, k-1} )^ 6\right),\\   
c^{(4)}_{4 \, k} & = & \frac{2 c^{(4)}_{4 \, k-1}}{C^2}+ 4\left(\frac{ c^{(3)}_{1 \, k-1} }{2 C^ {1/2} c^{(2)}_{1 \, k-1}} \right)^4+ O\left((c^{(3)}_{1 \, k-1} )^ 6\right),\\   
c^{(5)}_{1 \, k} & = & \frac{2 c^{(5)}_{1 \, k-1}}{C^{5/2}}+ \frac{n+6}{2}\left(\frac{ c^{(3)}_{1 \, k-1} }{2 C^ {1/2} c^{(2)}_{1 \, k-1}} \right)^5+ O\left((c^{(3)}_{1 \, k-1} )^ 7\right),\\   
c^{(5)}_{2 \, k} & = & \frac{2 c^{(5)}_{2 \, k-1}}{C^{5/2}}- 40\left(\frac{ c^{(3)}_{1 \, k-1} }{2 C^ {1/2} c^{(2)}_{1 \, k-1}} \right)^5+ O\left((c^{(3)}_{1 \, k-1} )^ 7\right),\\   
c^{(5)}_{3 \, k} & = & \frac{2 c^{(5)}_{3 \, k-1}}{C^{5/2}}+ 30\left(\frac{ c^{(3)}_{1 \, k-1} }{2 C^ {1/2} c^{(2)}_{1 \, k-1}} \right)^5+ O\left((c^{(3)}_{1 \, k-1} )^ 7\right),\\   
c^{(5)}_{4 \, k} & = & \frac{2 c^{(5)}_{4 \, k-1}}{C^{5/2}}+5\left(\frac{ c^{(3)}_{1 \, k-1} }{2 C^ {1/2} c^{(2)}_{1 \, k-1}} \right)^5+ O\left((c^{(3)}_{1 \, k-1} )^ 7\right). \label{111}
\eea
\ew
%It is important to observe from Eqs. (\ref{110}) - (\ref{111}) that the smallness condition for $ c^{(4)}_{l \, k} , \,  c^{(5)}_{l \, k}\, l=1,\ldots,4$ is satisfied only if an additional condition on $\sigma$ holds. This can be shown by considering  (\ref{110}) - (\ref{111})  at the fixed-point, and the definition (\ref{120}). For  example, the fixed-point value $c^{(4)}_{2 \, \ast}$ is given by 
%\[
%c^{(4)}_{2 \, \ast} = \frac{1}{1-\frac{2}{C^2}} \left[ - \frac{3}{2}\left(\frac{ c^{(3)}_{1 \, \ast} }{2 C^ {1/2} c^{(2)}_{1 \, \ast}} \right)^4+ O\left((c^{(3)}_{1 \, \ast} )^ 6\right) \right].
%\]
%It follows that $c^{(4)}_{2 \, \ast}$ is small only if 
%\be \label{121}
%\epsilon \ll \frac{1}{12}.
%\ee
 %It is easy to show that the condition (\ref{121}) holds at any perturbative order $p$. Eq (\ref{121}) physically represents the limit of validity of  the $\epsilon$ expansion for dimensions lower than the critical dimension, and corresponds to the value $\epsilon=1$ in the Ising model case, where $\epsilon \equiv 4-d$. As we will show in the following, the condition (\ref{121}) can be interpreted as a rough `smallness condition' for $\epsilon$ for the series of $\nu$ to be well-behaved.  \\

By looking at Eqs. (\ref{110}) - (\ref{111}) and using the definition (\ref{120}), it is easy to see that the coefficients $c^{(4)}_{l \, k}, \, c^{(5)}_{l \, k}$ scale to zero as $k\rightarrow \infty$ if $\epsilon < 1/12$. It is easy to find out that this is actually true for all the coefficients $c^{(j)}_{l \, k}$ with $j > 3$. Such a critical value $\epsilon=1/12$ will be reproduced also in the field-theoretical approach in Section \ref{zinn}.\smallskip \\

The evolution Eqs. (\ref{220}) - (\ref{111})  depend smoothly on the replica number $n$, so that the analytical continuation $n\rightarrow0$, can be done directly. By linearising the transformation (\ref{220}) - (\ref{111})   around the critical fixed-point $\{c^{(j)}_{l\, *}\}_{j,l}$ and computing the matrix $\mathcal{M}$, one can extract $\Lambda$, and so $\nu$ for $n=0$ to the order $\epsilon^2$ by using Eq. (\ref{80}). We find
\begin{eqnarray}\label{5}
\nu &= & 3 + 36 \epsilon +  \big[ 432 - 27 \big(50 + 55 \cdot 2^{1/3} + \\ \nonumber
& & + 53 \cdot  2^{2/3}\big) \log 2 \big] \epsilon^2 + O\left( \epsilon^3 \right).
\end{eqnarray}
%We observe that the second term of the expansion (\ref{5}) is much smaller than the first one if the condition (\ref{121}) is satisfied. 
\medskip \\

The one-loop result for $\nu$ is the same as that of the power-law interaction spin-glass  \cite{8} (where $\epsilon\equiv 3(\sigma-2/3)$). Notwithstanding this, the coefficients of the expansion in these two models will be in general different at two or more loops. As a matter of fact, the binary tree-structure of the interaction of the HEA emerges in the non-trivial $\log 2, \, 2^ {1/3}$ factors in the coefficient of $\epsilon^2$ in (\ref{5}), that come from the binary structure of the hierarchical tree and can't be there in the power-law case.
% Hence, the novelty of the HEA is that it yields a new series in $\epsilon$  can be evaluated much more easily than that of the long-range    \cite{8} and short-range   \cite{chen} spin-%glass models, thanks to the hierarchical symmetry structure making the RG equations simple. 
\medskip \\
Before discussing the result in Eq. (\ref{5}),  we point out that  Wilson's method explicitly implements the binary-tree structure of the model when approaching the IR limit. As a matter of fact, the hierarchical structure of the model  is explicitly exploited to construct the steps of the RG transformation. Nevertheless, if the IR limit is unique and well-defined, physical observables like $\nu$ must not depend on the technique we use to compute them in such a limit. It is thus important to verify that  Eq. (\ref{5})  does not depend on the method we used to reproduce the IR behavior of the theory. This has been done by reproducing Eq. (\ref{5}) with a quite different field-theoretical approach.

\bw
\begin{centering}

\begin{table}
\begin{centering}
\begin{tabular}{|c|c|c|c|c|}
\hline 
$j$ & \multicolumn{4}{c|}{$I_{l}^{(j)}[Q]$}\tabularnewline
\hline 
$2$ & \multicolumn{4}{c|}{$\text{Tr}[Q^{2}]$ } \tabularnewline
\hline 
$3$ & \multicolumn{4}{c|}{$\text{Tr}[Q^{3}]$} \tabularnewline
\hline 
$4$ & $\text{Tr}[Q^{4}]$ & $\text{Tr}[Q^{2}]^{2}$ & $\sum_{a\neq  c}Q_{ab}^{2}Q_{bc}^{2}$ & $\sum_{a b}Q_{ab}^{4}$\tabularnewline
\hline 
$5$ & $\text{Tr}[Q^{5}]$ & $\text{Tr}[Q^{2}]\text{Tr}[Q^{3}]$ & $\sum_{a b c d}Q_{ab}^{2}Q_{bc}Q_{bd}Q_{cd}$ & $\sum_{a b c}Q_{ab}^{3}Q_{ac}Q_{bc}$\tabularnewline
\hline
\end{tabular}
\par\end{centering}

\caption{Invariants generated to the order $p=5$. In each line of the table we show  the invariants $I^ {(j)}_1[Q],\ldots,I^ {(j)}_{n_j}[Q]$ from left to right.}
\label{tab1}
\end{table}
\end{centering}
\ew
\section{Field-theoretical method}\label{zinn}

Here the IR limit  is  performed by constructing a functional integral field theory and by removing its IR divergences within the minimal subtraction scheme. \\

While in Wilson's method  the IR limit was reached by  looking at the scale invariant fixed-points of  the recursion relation (\ref{2}) for $k\rightarrow \infty$, in this case the we will take before the large-$k$ limit, remove the resulting IR singularities through re-normalization, and then perform the scale-invariant limit by means of the Callan-Symanzik equation. \\

This computation is better performed by slightly changing the definition of the model. Indeed, the following re-definition of the interaction term  in Eq. (\ref{1})  
\begin{equation}\label{8}
\sum_{i<j}^{0,2^{k+1}-1} J_{ij} S_i  S_j \rightarrow \sum_{i=0}^{2^{k}-1} \sum_{j=2^{k}}^{2^{2k}-1} J_{ij} S_i  S_j. 
\end{equation}
is equivalent to the original definition (\ref{1}) and makes the field theory computations simpler. The equivalence of (\ref{8}) with the original definition (\ref{1}) can be shown  \cite{franzjorg} by observing that the scaling of the spin coupling in the model defined by Eq. (\ref{8}) differs  from that in Eq. (\ref{1})  for a constant multiplicative factor, and thus that  the two options are equivalent, and must yield the same critical exponents.
% M 
% Giorgio qui non sono sicuro di questa frase che segue. 
 Notwithstanding this, the critical temperature of the model defined by Eq. (\ref{1}) and that of the model defined by Eq. (\ref{8}) will be different, since  the spin couplings $J_{ij}$  differ by a multiplicative factor, resulting in a redefinition of $\beta$. Physically speaking, the  original definition   (\ref{1}) is such that, when two subsystems of $2^k$ spins are coupled to form a system with $2^{k+1}$ spins, one introduces couplings $J_{ij}$ between the two subsystems \textit{and} additional  couplings between   the spins within each subsystem,  while in (\ref{8}) only couplings between the two subsystems are introduced. \\

By iterating  the recursion relation for the Hamiltonian, one has an explicit form for  $H_{k}^ J[S]$ of a system of $2^ k$ spins in the large-$k$ limit. Then, the average of the replicated partition function is be expressed as an integral over the local overlap field  $Q_{i\, ab}\equiv S_i^a S_i^b$
\beas
 \mathbbm{E}_{J}\left[Z^n\right] &=& \\  \no
 \mathbbm{E}_J\left[ \sum_{ \{S_i^a\}_{i,a}} \exp\left(-\beta \sum_{a=1}^ {n} H_k^ J[S_0^ a,\ldots, S_{2^ k-1}^ a]\right)\right] &=&\\ \no 
 \int \left[ d Q \right] e^{-S[Q]}.&&
\eeas
By using dimensional analysis, it is easy to pick up the terms in $S[Q]$ that are relevant in the IR-limit. It is easy to check that $S[Q]$ is given by the sum of a quadratic term in $Q_{i\, ab}$, plus a cubic term, plus higher-degree terms. The dimensions of the field $Q_{i\, ab}$ can be computed by imposing the adimensionality of the quadratic term, and so the dimension of the coefficient $g$ of the cubic term and of those of the higher-degree terms. One finds that the dimensions of $g$ in energy is $[g]=3 \epsilon$. Thus, as in Wilson's method, the cubic term scales to zero in the IR limit for $\epsilon<0$, while a non-trivial fixed point appears for $\epsilon>0$. 
As in Wilson's method, it is easy to see that for $\epsilon<1/12$ all the higher-degree terms in $S[Q]$ scale to zero in the IR limit. Thus, the IR-dominant part of the action reads
\begin{eqnarray}\label{6}
S[Q] =\frac{1}{2} \sum_{i,j=0}^{2^k-1} \Delta_{ij} \text{Tr}\big[Q_i  Q_j \big]  +\frac{g}{3 !} \sum_{i=0}^{2^k-1}  \text{Tr}[Q_i^3],
\end{eqnarray}
where  and  $m \propto T-T_c$. 
The bare propagator $\Delta_{ij}$ actually depends on $i,j$  through the difference $\mathcal{I}(i)-\mathcal{I}(j)$, where the function
$\mathcal{I}(i)$ is defined as follows: given $i\in [0, 2^k -1]$ and its expression in base $2$
\begin{equation}
i = \sum_{j=0}^{k-1} a_{j} 2^j,\, \mathcal{I}(i) \equiv \sum_{j=0}^{k-1} a_{k-1-j} 2^j. 
\end{equation}
 Hence, the quadratic term of Eq. (\ref{6})  is not invariant under spatial translations. This would make any explicit computation of the loop-integrals, and so of the critical exponents, extremely  difficult to perform. This problem can be overcome by a re-labelling of the sites of the lattice   \cite{parisisourlas, meurice}
\[
\mathcal{I}(i) \rightarrow i, \, i=0, \ldots, 2^k-1.
\] 

 After relabelling one obtains that $\Delta_{ij}$ depends on $i,j$ just through the difference $i-j$,  thus $S[Q]$ is traslationally invariant, and the ordinary Fourier transform  techniques   \cite{fourier, meurice} can be employed. In particular, the Fourier representation of the propagator is 
\be  \label{prop}
\Delta_{ij} = \frac{1}{2^ k} \sum_{p=0}^ {2^k-1} \exp\left[\frac{-2 \pi \imath p(i-j)}{2^ k}\right] \left( |p|_2^{2 \sigma-1}+ m \right). 
\ee 
where $|p|_2$  is the di-adic norm of $p$  \cite{parisisourlas}, and the mass $m$ has dimension $[m]=2\sigma -1$. \smallskip\\
An interesting feature of the action (\ref{6}) is the fact the propagator $\Delta$ in Eq. (\ref{prop}) depends on the momentum $p$ through its diadic norm $|p|_2$.  If we look at the original derivation of the recursion RG equation
for the Ising model in finite dimension (in particular to the Poliakoff
derivation  \cite{wilsonkogut}), we find that the basic
approximation was to introduce an ultrametric structure in momentum
space: the momentum space is divided in shells an the sum of two
momenta of a given shell cannot give a momentum of a higher momentum
scale cell. This has a nice similarity with the metric properties of the diadic norm, where if $p_1,\, p_2$ are two integers, their diadic norms satisfies  \cite{parisisourlas}
$|p_1-p_2|_2 \leq \max \left( |p_1|_2,|p_2|_2\right)$. \medskip\\

The field theory defined by Eq. (\ref{6}) reproduces the  $\text{Tr}[Q^3]$ interaction term of the well-know effective actions describing the spin-glass transition in short-range   \cite{chen} and long-range   \cite{mau, 8} spin-glasses. Notwithstanding this similarity, the novelty of the HEA is that a high-order $\epsilon$-expansion can be quiet easily  automatized within Wilson's method, by means of a symbolic manipulation program solving the simple RG equation (\ref{2}) to high orders in $\epsilon$. This is not true for such short and long-range   \cite{chen, mau,8} models, where the only approach to compute the exponents is the field-theoretical one. Indeed, nobody ever managed to automatize at high orders a computation of the critical exponents within the field-theoretical minimal subtraction scheme, either for the simplest case of the Ising model, because such an automatization to high orders is not an easy task  \cite{GR}. \\ 
% M
% Qui tu mi avevi detto che c'era qualcuno che aveva quasi automatizzato il conto per Isinig, ma non riesco a leggerne il nome dal foglio, né a trovarlo nella letteratura... Potresti aggiungere tu una frase in merito e la referenza? 

\begin{figure}
\vspace{1cm}
\centering
%\resizebox{1\linewidth}{!}{\input{graph.1}}
\includegraphics[width=2cm]{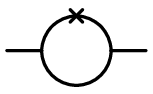}\hspace{.5cm}
\includegraphics[width=2cm]{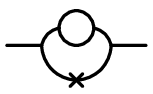}\hspace{.5cm}
\includegraphics[width=2cm]{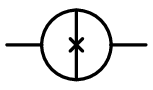} \vspace{4mm}\\
\includegraphics[width=2cm]{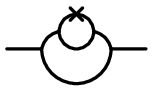}\hspace{.5cm}
\includegraphics[width=2cm]{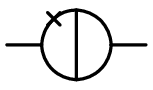}\hspace{.5cm}
\includegraphics[width=2cm]{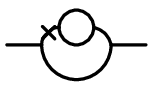}

\caption{One and two-loop Feynman diagrams contributing to $\Gamma_r^{(2,1)}$. The crosses represent $\text{Tr}\left[Q^2\right]$ insertions. From left to right, such diagrams computed at zero external momenta are equal to $\mathscr{I}_1,\mathscr{I}_2,\mathscr{I}_3,\mathscr{I}_4,\mathscr{I}_5,\mathscr{I}_6$ respectively. }
\label{fig1}
\end{figure}

\begin{figure} 
\vspace{5mm}
\centering
%\resizebox{1\linewidth}{!}{\input{graph.1}}
\includegraphics[width=2cm]{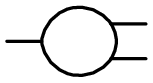}\hspace{.5cm}
\includegraphics[width=2cm]{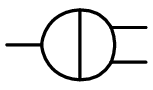}\hspace{.5cm}
\includegraphics[width=2cm]{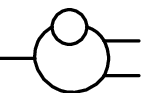}\\ \vspace{5mm}
\includegraphics[width=2cm]{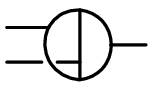}
\caption{One and two-loop Feynman diagrams contributing to $\Gamma_r^{(3,0)}$.  From left to right, such diagrams computed at zero external momenta are equal to $\mathscr{I}_7,\mathscr{I}_8,\mathscr{I}_9,\mathscr{I}_{10}$ respectively. The last diagram is non-planar.}
\label{fig2}
\end{figure}

The additional order of difficulty in computing $\nu$ with the field-theoretical approach for the HEA will be clear in the following, as we sketch out its main steps. \bigskip \\
 The field theory defined by Eq. (\ref{6}) can be now analysed within the loop expansion framework. The renormalized mass and coupling constant are defined as 
 \bea
 m&= &m_r+\delta m ,\\ 
 g&=&m_r^{\frac{3 \epsilon}{2 \sigma -1 }}g_r Z_g. \label{226}
 \eea
  We define the  one-particle-irreducible  \cite{zinnjustin} ($1$PI) renormalized correlation functions
  \bw
\begin{eqnarray*}
 \Gamma^{(m,l)}_{r}(a_1 b_1 i_1 \cdots a_m b_m i_m; j_1 \cdots j_l; g_r, m_r^ {\frac{1}{ 2\sigma-1}}) &  \equiv  &  Z_2^ l    \Gamma^{(m,l)}(a_1 b_1 i_1 \cdots a_m b_m i_m; j_1 \cdots j_l;g,m^{\frac{1}{ 2\sigma-1}}),  \\ \nonumber
  \Gamma^{(m,l)}(a_1 b_1 i_1 \cdots a_m b_m i_m; j_1 \cdots j_l;g,m^{\frac{1}{ 2 \sigma -1}}) &\equiv &  2^{-l} \left\langle Q_{i_1\,a_1 b_1} \cdots Q_{i_m\, a_m b_m} \text{Tr}\left[ Q_{j_1}^2\right] \cdots \text{Tr}\left[ Q_{j_l}^2\right] \right\rangle_{1\text{PI}},
\end{eqnarray*}
\ew
in terms of the renormalized  parameters $m_r,g_r$. Since this model has long-range interactions, the field $Q_{ab}$ is not renormalized, and  \cite{zinnjustin} $Z_Q=1$. Hence, all we need to compute $\nu$, are   \cite{zinnjustin} the renormalization constants $Z_g,Z_2$ and $\delta m$. This can be done by computing the IR-divergent parts of $\Gamma_r^{(3,0)},  \Gamma_r^{(2,1)}$  with the minimal subtraction scheme  \cite{zinnjustin}. In other words, one takes the IR-limit $m_r\rightarrow 0$, and systematically removes the resulting $\epsilon$-singular parts of the correlations functions, by absorbing them into the renormalization constants $Z_g,Z_2$. \\
 The  Feynmann diagrams contributing to  $\Gamma_r^{(3,0)},  \Gamma_r^{(2,1)}$ are shown in Figs. \ref{fig1}, \ref{fig2}, and their   singular parts are in the form of $1/\epsilon, 1/\epsilon^2$-poles. \\

Here we show by a simple example how the $\epsilon$-divergent part of such diagrams can be computed. Let's consider the one-loop expansion of $\Gamma^{(3,0)}_r$. This is obtained by picking up the $\text{Tr}[\mathcal{Q}^ 3]$-term in the renormalized $1$PI generating functional  \cite{zinnjustin}
\bea  \label{115}
\Gamma_r[\mathcal{Q}] & = & \frac{1}{2} \sum_{i j} \Delta_{ij} \text{Tr}[\mathcal{Q}_i \mathcal{Q}_j ] + \\ \no 
&&\frac{m_r^ {3 \epsilon} g_r}{3!} \sum_i \text{Tr}[\mathcal{Q}_i^3] \left[ Z_g + \frac{n-2}{8} \mathscr{I}_1 g_r^2\right] + \\ \no 
&&  + O(g_r^ 5).
\eea
The loop-integral
\bea \label{112}
\mathscr{I}_1 &\equiv &  \frac{1}{2^ k} \sum_{p=0}^ {2^ k-1} \frac{1}{\left(m_r +\delta m+ |p|_2^ {2 \sigma-1}\right)^ 3} 
\eea
is represented by the  first diagram in Fig. \ref{fig2}. \\
Eq. (\ref{112}) has a well-defined limit for $k\rightarrow \infty$. Indeed, thanks to the translational invariance of the theory,  the argument of the sum in the right side of (\ref{112}) depends on $p$ just through its diadic norm. It follows that the sum $\mathscr{I}_1$ can be transformed into a sum over all the possible values of $|p|_2$. Indeed, using the standard result   \cite{parisisourlas} that the number of integers $p \in [0, 2^ k -1] $ such that $|p|_2=2^ {-j}$, i. e. the volume of the diadic shell, is given by 
$
2^ {-j+k-1},
$
Eq. (\ref{112}) becomes 
\bea \label{113}
\mathscr{I}_1 & = & \sum_{j=0}^ {k-1} 2^ {-j-1}  \frac{1}{\left[m_r   +\delta m+ 2^ {-j(2 \sigma-1)}\right]^ 3}\\ \no 
& \rightarrow & \sum_{j=0}^ {\infty} 2^ {-j-1}  \frac{1}{\left[m_r  +\delta m+ 2^ {-j(2 \sigma-1)}\right]^ 3},
\eea
where in the second line of Eq. (\ref{113}) the $k \rightarrow \infty$ limit has been taken, since the series in the first line is convergent. 
By using the fact that $\delta m = O(g_r^2)$, we can rewrite (\ref{113}) as 
\bea \label{218}
\mathscr{I}_1 & = & \sum_{j=0}^ {\infty} 2^ {-j-1}  \frac{1}{\left[m_r   +   2^ {-j(2 \sigma-1)}\right]^ 3}+\\ \no
&&+ O(g_r^2).
\eea
It easy to see that $\mathscr{I}_1$ is IR-divergent for $m_r \rightarrow 0$. 
 Indeed, in the limit $m_r \rightarrow 0$ the sum over $j$ in Eq. (\ref{218}) is dominated by the terms in the IR region $2^{-j}=|p|_2\rightarrow 0$. The $j$s corresponding to this region are given by the relation $m_r \approx 2^ {-j}$, and go to infinity as $m_r \rightarrow 0$, yielding a divergent sum in $\mathscr{I}_1$.\\
   In the IR region, the sum in the right side of Eq. (\ref{218}) can be approximated by an integral, since the integrand function is almost constant in the interval $[ j,j+1]$ for large $j$. Setting $q\equiv 2^ {-j}$,  for $m_r \rightarrow 0$ we have $- q   \log 2 \, dj = dq$, and

\bea \label{119}
\mathscr{I}_1 & = & \frac{1}{2 \log 2}  \int_0 ^ 1  \frac{dq}{\left[m_r  + q^ {2 \sigma-1}\right]^ 3} +\\ \no
&& + O(g_r^2)\\ \no 
& = & \frac{m_r^ {-\frac{6 \epsilon}{2\sigma -1}}}{2 \log 2}  \int_0 ^ {m_r^{-\frac{1}{2 \sigma -1}}}  \frac{dx}{\left(1+ x^ {2 \sigma-1}\right)^ 3} + O(g_r^2)\\ \no 
& \rightarrow & \frac{m_r^ {-\frac{6 \epsilon}{2\sigma -1}}}{2 \log 2}  \int_0 ^ {\infty}  \frac{dx}{\left(1+ x^ {2 \sigma-1}\right)^ 3} + O(g_r^2).
\eea
The integral in the right side of the last line in Eq. (\ref{119}) is convergent for $\epsilon>0$, and diverges as $\epsilon \rightarrow 0$. Its $\epsilon$-divergent part can be easily evaluated
\bea \label{114}
\mathscr{I}_1 & = &  \frac{m_r^ {-\frac{6 \epsilon}{2\sigma -1}}}{4 \log 2}  \Gamma\left( 3 + \frac{1}{1-2 \sigma}\right) \times \\ \no 
&& \times \Gamma\left(  1 + \frac{1}{1-2 \sigma} \right) + O(g_r^2)\\ \no 
& = & m_r^ {-\frac{6 \epsilon}{2\sigma -1}} \left[\frac{1}{12\epsilon \log 2  } + O_{\epsilon}(1) \right]+ O(g_r^2),
\eea
where $\Gamma$ is the Euler-Gamma function and $O_{\epsilon}(1)$ denotes terms that stay finite as $\epsilon \rightarrow 0$. As we will show in the following, these finite terms will give a contribution to the renormalization constants at two loops. \\
By plugging Eq. (\ref{114}) into Eq. (\ref{115}), one can compute the $g_r^2$-coefficient of $Z_g$ by imposing that the $\epsilon$-singular part of $\mathscr{I}_1$ is cancelled by $Z_g$. For $n=0$ we have
\be \label{221}
Z_g = 1 + \frac{1}{48   \epsilon \log 2}g_r^2 + O(g_r^ 4).
\ee 
By repeating the same computation for $\Gamma_r[\mathcal{Q},K] $, and imposing that the $\sum_{i=0}^{2^ k-1}K_i \mathcal{Q}_i^2$-term is finite, i. e. that $\Gamma_r^{(2,1)}$ is finite, we obtain
\[
Z_2 = 1 + \frac{1}{24   \epsilon \log 2}g_r^2 + O(g_r^ 4).
\]
\bigskip \\
%
%An explicit evaluation of such loop integrals shows that IR divergences can be re-absorbed into the re-normalization constants  $Z_g,\, Z_2$ by means of the minimal subtraction scheme   \cite{zinnjustin}. 
Such procedure has been pushed to two loops by an explicit calculation. Even if the evaluation of the $\epsilon$-divergent part of the two-loop diagrams is more involved, the techniques and underlying ideas are exactly the same as those used to compute the one-loop diagram $\mathscr{I}_1$. In Figs. \ref{fig1}, \ref{fig2} we show the Feynman diagrams contributing to the finiteness conditions of  $\Gamma^{(2,1)}_{r}$ and of $\Gamma^{(3,0)}_{r}$ respectively. The diagrams in Fig. \ref{fig1} evaluated at zero external momenta will be denoted by $\mathscr{I}_1,\ldots,\mathscr{I}_6$, while those in Fig. \ref{fig2} by $\mathscr{I}_7,\ldots,\mathscr{I}_{10}$. It is easy to show that the equalities 
\beas
\mathscr{I}_1&=&\mathscr{I}_7,\\
 \mathscr{I}_2&=&\mathscr{I}_6=\mathscr{I}_9,\\
   \mathscr{I}_3&=&\mathscr{I}_{10} ,\\
    \mathscr{I}_4&=&\mathscr{I}_5=\mathscr{I}_8
\eeas
hold, and so that all we need to compute the  renormalization constants are $\mathscr{I}_1,\mathscr{I}_2,\mathscr{I}_3,\mathscr{I}_4$. 
$\mathscr{I}_1$ is given by Eq. (\ref{112}), while the other loop integrals are 
\bw
\beas
\mathscr{I}_2 & = & \frac{1}{2^ {2k}} \sum_{p=0}^ {2^ k-1}  \sum_{q=0}^ {2^ k-1}  \frac{1}{\left(m_r +\delta m+ |p|_2^ {2 \sigma-1}\right)^ 4\left(m_r +\delta m+ |q|_2^ {2 \sigma-1}\right) \left(m_r +\delta m+ |p-q|_2^ {2 \sigma-1}\right)} \\
\mathscr{I}_3 & = & \frac{1}{2^ {2k}} \sum_{p=0}^ {2^ k-1}  \sum_{q=0}^ {2^ k-1}  \frac{1}{\left(m_r +\delta m+ |p|_2^ {2 \sigma-1}\right)^ 2\left(m_r +\delta m+ |q|_2^ {2 \sigma-1}\right) ^2\left(m_r +\delta m+ |p-q|_2^ {2 \sigma-1}\right)^2} \\
\mathscr{I}_4 & = & \frac{1}{2^ {2k}} \sum_{p=0}^ {2^ k-1}  \sum_{q=0}^ {2^ k-1}  \frac{1}{\left(m_r +\delta m+ |p|_2^ {2 \sigma-1}\right)^ 3\left(m_r +\delta m+ |q|_2^ {2 \sigma-1}\right) ^2\left(m_r +\delta m+ |p-q|_2^ {2 \sigma-1}\right)}.
\eeas
\ew

 In the limit $m_r \rightarrow 0$,  $\mathscr{I}_2,\mathscr{I}_3,\mathscr{I}_4$ are given by 
\bw
\beas
\mathscr{I}_2 & = & m_r^ {\frac{-12 \epsilon}{2\sigma -1}}  \left[ \left( \frac{2}{2^ {2/3}-1}- \frac{1}{2^ {1/3}-1} -1\right)\frac{1}{48 \epsilon \log 2}  \right]+O_{\epsilon}(1),\\ 
\mathscr{I}_3 & = & m_r^ {\frac{-12 \epsilon}{2\sigma -1}}    \left(  \frac{2}{2^ {1/3}-1}\frac{1}{16 \epsilon \log 2}  \right)+O_{\epsilon}(1),\\ 
\mathscr{I}_4 & = & m_r^ {\frac{-12 \epsilon}{2\sigma -1}}   \left\{ \frac{1}{2}  \left[ \frac{1}{(12 \epsilon \log 2)^2} - \left( \frac{3}{8 (\log 2)^2} + \frac{1}{48 \log 2} \right)\frac{1}{\epsilon} \right] +\left( \frac{1}{2^ {1/3}-1} + \frac{1}{2^ {2/3}-1} \right) \frac{1}{48 \epsilon \log 2}  \right\}+O_{\epsilon}(1). 
\eeas
\ew 
The finiteness of $\Gamma^{(3,0)}_{r}$ is imposed by making finite the $\sum_{i=0}^ {2^ k-1}\text{Tr}[\mathcal{Q}_i^3]$-term in the $1$PI generating functional $\Gamma_r[\mathcal{Q}]$. The finiteness  of $\Gamma^{(2,1)}_{r}$ is imposed  by making finite the $\sum_{i=0}^{2^ k-1}K_i \mathcal{Q}_i^2$-term in the $1$PI the generating functional with $\text{Tr}[\mathcal{Q}_i^2]$-insertions, $\Gamma_r[\mathcal{Q},K]$. The two-loop expansion of $\Gamma_r[\mathcal{Q}]$ and of $\Gamma_r[\mathcal{Q},K]$ read
\bw
\bea \label{215}
\Gamma_r[\mathcal{Q},0] & = & g_r \sum_{i=0}^ {2^ k-1} \text{Tr}[\mathcal{Q}_i^3] \Bigg\{ \frac{m_r^ {\frac{3 \epsilon}{2 \sigma-1}}Z_g}{3 ! } + \frac{g_r^2(n-2)}{6}\left( \frac{m_r^ {\frac{3 \epsilon}{2 \sigma-1}} Z_g}{2} \right)^ 3 \mathscr{I}_1 + \\ \no 
&& + \frac{g_r^4 m_r^ {\frac{15 \epsilon}{2 \sigma-1}} Z_g^ 5} {3! 2^ 7} \left[6(n-2)^2  \mathscr{I}_8 + 6(n-2)^ 2 \mathscr{I}_7 + 2 (n(n-1)-4-(n-2)^2)\mathscr{I}_{10}\right] + O(g_r^ 6)
\Bigg\} + \cdots,\\  \label{216}
\Gamma_r[\mathcal{Q},K] & = & \sum_{i=0}^ {2^ k-1} K_i \text{Tr}[\mathcal{Q}_i^2] \Bigg\{ \frac{Z_2}{4} + \frac{g_r^2 Z_2 Z_g^2 m_r^{\frac{6 \epsilon}{2 \sigma-1}}(n-2)}{16} \mathscr{I}_7 + \frac{g_r^ 4 m_r^ {\frac{12 \epsilon}{2 \sigma-1}}(n-2)^2}{2^ 8 }\small[ 2 \left( 3 \mathscr{I}_2 + 2 \mathscr{I}_4  \right)  + 4  \mathscr{I}_5 +\\ \no 
&&+ \mathscr{I}_3 \small]+ O(g_r^ 6)\Bigg\} + \cdots,
\eea
\ew
where the $\cdots$ in Eq. (\ref{215}) stands for terms that are not cubic in $\mathcal{Q}_i$, while the $\cdots$ in Eq. (\ref{216}) for terms that are not quadratic in $\mathcal{Q}_i$ and linear in $K_i$. The renormalization constants $Z_g,\,Z_2$ are calculated by imposing that the renormalized correlation functions $\Gamma^{(3,0)}_{r}$, $\Gamma^{(2,1)}_{r}$ are finite, i. e. that the term in curly brackets in Eq. (\ref{215}) and that in Eq. (\ref{216}) have no singularities  \cite{zinnjustin} in $\epsilon$.  At this purpose, we observe that the finite part of the integral $\mathscr{I}_1$  contributes to the renormalization constants at two loops. For example, let us consider the second addend in curly brackets in the right side of Eq. (\ref{215}). By using Eq. (\ref{114}) and the one-loop result (\ref{221}) for $Z_g$, it is easy to see that this term  produces an $\epsilon$-divergent  term, given by 
\be \label{225}
 \frac{g_r^2(n-2)}{48}          m_r^ {\frac{3 \epsilon}{2 \sigma-1}}    \frac{3}{48   \epsilon \log 2}g_r^2 O_{\epsilon}(1),
\ee
where $O_{\epsilon}(1)$ is the finite part of $\mathscr{I}_1$ in Eq. (\ref{114}). The term in  Eq. (\ref{225}) is of $O(g_r^4)$ and singular in $\epsilon$. Hence, it contributes to the $O(g_r^4)$-term in $Z_g$. \medskip \\
After setting $n=0$ and imposing the finiteness conditions, we find
\bw
\bea
Z_g &=&1+\frac{g_r^2}{48 \epsilon \log 2}+g_r^4 \left[ \frac{1}{1536 \epsilon^2 (\log 2)^2}+\frac{5+2 \cdot 2^{2/3}}{512 \epsilon \log 2}\right]+O(g_r^6), \label{228}\\ 
Z_2&=&1 + \frac{g_r^2}{24 \epsilon \log 2}+ g_r^4 \left[ \frac{1}{576 \epsilon^2 (\log 2)^2} - 5\frac{ (1 + 11 \cdot 2^{1/3} + 7 \cdot 2^{2/3})}{ 2304 \epsilon  \log 2}\right] +O(g_r^ 6). \label{229}
\eea
\ew
It is also easy to verify that  $\delta m = O(g_r^ 4)$.\\

Once the  IR-safe renormalized theory has been constructed,  the effective coupling constant $g(\lambda)$ at the energy scale  $\lambda$ is computed form the Callan-Symanzik equation in terms of the $\beta$-function by setting $\mu \equiv m_r ^{\frac{1}{2\sigma-1}}$
\be 
\beta(g_r) = \left. \mu \frac{\partial g_r}{\partial \mu} \right | _{g,m},\,
\beta(g(\lambda)) = \lambda \frac{d g(\lambda)}{d \lambda}.
\ee 
$\beta(g_r)$ can be explicitly computed in terms of the renormalization constant $Z_g$ by applying $\left. \mu \frac{\partial }{\partial \mu} \right | _{g,m}$ on both sides of Eq. (\ref{226})
\be \label{227}
0 =\left. \mu \frac{\partial }{\partial \mu} \right | _{g,m} \left( \mu ^{3 \epsilon}g_r Z_g \right).
\ee
The right side of Eq. (\ref{227}) can be then worked out explicitly by using the two-loop result (\ref{229}) and substituting systematically $\beta(g_r)$ to  $\left. \mu \frac{\partial g_r}{\partial \mu} \right | _{g,m}$. In this way, an explicit equation for $\beta(g_r)$ is obtained. By solving perturbatively this equation to $O(g_r^ 5)$ one finds
\be \label{beta}
\beta(g_r)= -3 \epsilon g_r + \frac{g_r^ 3}{8 \log 2} + 3 \frac{5+2 \cdot 2^{2/3}}{128 \log 2} g_r^ 5 + O(g_r^ 7).
\ee
Setting $g_r^\ast \equiv g(\lambda=0)$, we see from Eq.  (\ref{beta}) that the fixed point $g_r^\ast=0$ is stable only for $\epsilon<0$, while for $\epsilon>0$ a non-Gaussian fixed point $g_r^{*}$ of order $\epsilon$ arises, as predicted by dimensional considerations and by Wilson's method.
Now the IR-limit $\lambda\rightarrow 0$ can be safely taken, and the scaling relations yield $\nu$  in terms of $g_r^\ast $ and $Z_2$ 
\begin{equation} \label{7}
\eta_2[g_r]  \equiv \mu  \left. \frac{\partial \log Z_2}{\partial \mu } \right|_{g,\, m}, \, \nu= \frac{1}{\eta_2[g_r^{*}] + 2\sigma -1}. 
\end{equation}
By plugging the two-loop result for $g_r^{*}$ and  $Z_2$ into Eq. (\ref{7}), we reproduce  the result (\ref{5}) derived within Wilson's method.\medskip \\
We observe that the analytical effort to derive the coefficients of the $\epsilon$-expansion in this field-theoretical approach is much bigger than that of Wilson's method. Indeed, in the minimal  subtraction scheme,  additional calculations are needed to extract the coefficients of the  poles in $1/\epsilon, \, 1/\epsilon^2$ of the Feynamnn diagrams in Figs. \ref{fig1}, \ref{fig2}.
%CCCCCCC
 It follows that for an automatized implementation of the $\epsilon$-expansion to high orders, Wilson's method turns out to be much better-performing that the field theoretical method.
%CCCCCCC
 Notwithstanding this, the tensorial operations needed to compute $Q$-dependence of the diagrams in this field theoretical approach turn out to be exactly the same as those needed in Section \ref{wilson}, and  no additional effort has been required to compute them.

\section{Conclusions}\label{conc}
%CCCCCCCCCCCCCCCC
 In a previous work   \cite{io}, we  set up two perturbative approaches to compute the IR behavior of a strongly-frustrated non-mean field spin-glass system, the HEA model. The two methods  yield the same prediction at two loops for the critical exponent $\nu$ related to the divergence of the correlation length. \\
 
  In this work  the two-loop computation is shown in all its most relevant details, so that the reader can reproduce it. 
  Moreover, we show the underlying renormalization group ideas implemented in the  two computation methods.  Amongst these, the existence of a characteristic length $\xi$ diverging at the critical point, where the theory is invariant with respect to changes in its energy scale.  
  %In that regard, we show that the first method explicitly exploits the hierarchical structure of the model, and implements a Wilson-like coarse graining technique to reproduce the IR limit. The second method relies on the construction of an effective field theory reproducing the IR-limit by means of the Callan-Symanzik equation.
   %The two methods  yield the same prediction at two loops for the critical exponent $\nu$ related to the divergence of $\xi$. This shows that the IR-limit of the theory is well-defined and independent on the actual method one uses to reproduce it, and that an $\epsilon$-expansion can be consistently set up without ambiguities.  In both methods, we implemented the basic RG underlying ideas.
   In addition, we   show with an explicit example that such a computation of the critical exponents could be quiet easily automatized, i. e.  implemented in a computer program, in order to push this $\epsilon$-expansion to high orders in $\epsilon$, and so eventually make this theory physically predictive. Indeed, we give a graphical interpretation of the cumbersome tensorial operations needed  to compute $\nu$ and previously used in \cite{io}. Such a graphical method makes the calculations much more straightforward and elegant.
 We observe that once this high-order series in $\epsilon$ will be known, some resummation technique will be needed to make the theory predictive, because the series  has probably a non-convergent behavior. 
%, and universality.
%Thanks to the hierarchical symmetry of the model, a high-order development of the $\epsilon$-expansion could be quiet easily automatized by means of a symbolic manipulation program, by using  Wilson's method. Once the high-order coefficients of this series are known, one could explore the convergence and summability conditions of the $\epsilon$-expansion  \cite{zinnjustin}.
If the high-orders series could be made convergent by means of some appropriate re-summation technique, this calculation  would yield an analytical control on the critical exponents, resulting in a precise prediction for a non-mean field spin-glass mimicking a real system.   \smallskip\\
%CCCCCCCCCCCCCC

We  are glad to thank N. Sourlas, S. Franz and  M. M\'ezard   for  interesting discussions and suggestions.\\
 \\
\bibliographystyle{unsrt}

\begin{thebibliography}{10}
\bibitem{MPV}
M.~M\'ezard, G.~Parisi and  M.~A. Virasoro, \emph{Spin glass theory and beyond}, World Scientific, (1987).
%
\bibitem{castellani-cavagna}
T.~Castellani and A.~Cavagna, J. Stat. Mech. \textbf{P05012} (2005).
%
\bibitem{parisi}
G.~Parisi,  J. Phys. A: Math. Gen. \textbf{13} 1101 (1980).
%
\bibitem{6}
C.~De~Dominicis and I.~Giardina, \emph{Random fields and spin glasses: a field theory approach}, Springer, (2006).
%  
\bibitem{7}
J. H.~Chen and T. C.~Lubensky, Phys. Rev. B \textbf{16} (1977), 2106.
%
\bibitem{8}
G.~Kotliar, P.W.~Anderson and D.L.~Stein, Phys. Rev. B \textbf{27} (1983), 602.
%  
\bibitem{zinnjustin}
 J.~Zinn-Justin, Int.\ Ser.\ Monogr.\ Phys.\  {\bf 113}, 1 (2002).
  %
 \bibitem{wilsonkogut}
 K.~G.~Wilson and J.~B.~Kogut, Phys.\ Rept.\  {\bf 12}, 75 (1974).
 %
\bibitem{dyson}
F.J. Dyson, Comm. in Math. Phys. \textbf{12} (1969), 91.
%
\bibitem{collet-eckmann}
P.~Collet, J.~Eckmann, \emph{A renormalization group analysis of the hierarchical model in statistical mechanics}, Springer-Verlag, (1978).
 %
 \bibitem{cassandro}
 M.~Cassandro and G.~Jona-Lasinio,
 Adv.\ in\ Physics\ {\bf{27}}, 6 (1978).
 %
 \bibitem{theumann1}
A. Theumann, Phys. Rev. B 21 (1980) 2984.
%
\bibitem{theumann2}
A. Theumann, Phys. Rev. B 22 (1980) 5441.
% 
 \bibitem{franzjorg}
 S.~Franz, T.~J\"{o}rg and G.Parisi,
 J.\ Stat.\ Mec.,  P02002 (2009).
 %
 \bibitem{castellana}
 M.~Castellana, A.~Decelle, S.~Franz, M.~M\'ezard and G.~Parisi, Phys.\ Rev.\ Lett.\ {\bf{104}}, 12 (2010).
 %
 \bibitem{XXX} Y. V. Fyodorov, A. Ossipov, A. Rodriguez, J. Stat. Mech. (2009) L12001
 %
 \bibitem{parisisourlas}
 G.~Parisi and N.~Sourlas,
 Eur.\ Phys.\ Jour.\ B\ {\bf{14}}, 3 (2000). 
 %
 %
 \bibitem{meurice} 
 Y. Meurice, J.Math.Phys. 36 (1995) 1812-1824.
 %
  \bibitem{chen}
A. B. Harris, T. C. Lubensky and J. H. Chen, Phys. Rev. Lett. 36 (1976) 415.
%
\bibitem{mau}
M. Chung Chang and J. Sak, Phys. Rev. B 29 (1984) 2652.
 % 
  \bibitem{GR}
R. Guida, P.  Ribeca J.Stat.Mech.20096:P02007,2006
%
 \bibitem{fourier} 
 M. Taibleson,  {\em Fourier Analysis on Local Fields}, Princeton University Press (1976).
 %
 \bibitem{berker1}
 A. N. Berker, S. Ostlund, J. Phys. C: Solid State Phys. \textbf{12} (1979), 4961–4975.
%
\bibitem{berker2}
 S. R. McKay, A. N. Berker, S. Kirkpatrick, Phys. Rev. Lett. \textbf{48}f (1982), 767.
%
\bibitem{berker3}
M. Ohzeki, H. Nishimori,  A. N. Berker, Phys. Rev. E \textbf{77}, 061116 (2008).
%
\bibitem{berker4}
G. G\"{u}lpınar and A. N. Berker, Phys. Rev. E \textbf{79}, 021110 (2009).
%
\bibitem{berker5}
A. N. Berker, M. Hinczewski, and R. R. Netz, Phys. Rev. E \textbf{80}, 041118 (2009).
%
\bibitem{berker6}
A. N. Berker, Phys. Rev. E \textbf{81}, 043101 (2010).
%
\bibitem{gardner}
E. Gardner, J. Physique \textbf{45}, 1755 (1984).
%
\bibitem{io}
M. Castellana, G. Parisi, Phys. Rev. E \textbf{82}, 040105(R) (2010)
%
\bibitem{alignment}
B. D. McKay, \textit{Practical Graph Isomorphism, Congressus Numerantium}, \textbf{30} (1981), 45-87
%
\bibitem{young1}
H. G. Katzgraber and A. P. Young,  Phys. Rev. B \textbf{67} 134410 (2003).
%
\bibitem{young2}
H. G. Katzgraber and A. P. Young,  Phys. Rev. B \textbf{72} 184416 (2005).
%
\bibitem{young3}
H. G. Katzgraber, A. K.  Hartmann  and A. P. Young, 2008 Computer Simulation Studies in Condensed Matter
Physics XXI ed D P Landau, S P Lewis and H B Schuttler (Heidelberg: Springer)
%
\bibitem{leuzzi}
L. Leuzzi,  J. Phys. A: Math. Gen. \textbf{32} 1417 (1999)
%
\bibitem{parisi2}
L. Leuzzi, G. Parisi, F. Ricci-Tersenghi, and J. J. Ruiz-Lorenzo, Phys. Rev. Lett. \textbf{101}, 107203 (2008).



\end{thebibliography}

\end{document}